\documentclass{article}
\usepackage[utf8]{inputenc}
\usepackage{graphicx}
\usepackage{amsmath}
\usepackage{amssymb}
\usepackage[T1]{fontenc}
\usepackage{authblk}
\usepackage{xcolor}
\usepackage{soul}

\usepackage{cite}

\title{End-to-end metasurface inverse design for single-shot multi-channel imaging}
\author[1]{Zin~Lin\thanks{zinlin@mit.edu}}
\author[1]{Rapha{\"e}l~Pestourie}
\author[2]{Charles~Roques-Carmes}
\author[3]{Zhaoyi~Li}
\author[3]{Federico~Capasso}
\author[2,4]{Marin~Solja\v{c}i\'{c}}
\author[1]{Steven~G.~Johnson}
\affil[1]{Department of Mathematics, Massachusetts Institute of Technology, Cambridge MA 02138, USA}
\affil[2]{Research Lab of Electronics, Massachusetts Institute of Technology, Cambridge MA 02138, USA}
\affil[3]{John A. Paulson School of Engineering and Applied Sciences, Harvard University, Cambridge MA 02138, USA}
\affil[4]{Department of Physics, Massachusetts Institute of Technology, Cambridge MA 02138, USA}

\date{\today}

\begin{document}

\maketitle

\begin{abstract}
We introduce end-to-end metaoptics inverse design for multi-channel imaging: reconstruction of depth, spectral and polarization channels from a single-shot monochrome image. The proposed technique integrates a single-layer metasurface frontend with an efficient Tikhonov reconstruction backend, without any additional optics except a grayscale sensor. Our method yields multi-channel imaging by spontaneous demultiplexing: the metaoptics front-end separates different channels into distinct spatial domains whose locations on the sensor are optimally discovered by the inverse-design algorithm. We present large-area metasurface designs, compatible with standard lithography, for multi-spectral imaging, depth-spectral imaging, and ``all-in-one'' spectro-polarimetric-depth imaging with robust reconstruction performance ($\lesssim 10\%$ error with 1\% detector noise). In contrast to neural networks, our framework is physically interpretable and does not require large training sets. It can be used to reconstruct arbitrary three-dimensional scenes with full multi-wavelength spectra and polarization textures.
\end{abstract}

\newpage
\section{Introduction}
Metasurfaces have been heralded as a revolutionary platform for realizing complex functionalities and compact form factors inaccessible to conventional refractive or diffractive optics~\cite{yu2011light,khorasaninejad2017visible,chen2018broadband,engelberg2020advantages}.
Meanwhile, an emerging ``end-to-end'' paradigm in computational imaging, in which an optical frontend is optimized in conjunction with an image-processing backend, has received increasing attention due to successful applications in diffractive optics~\cite{sitzmann2018end,baek2020end}. More recently, the end-to-end paradigm has been introduced to full-wave vectorial nanophotonic and metasurface frontends~\cite{lin2021end,burgos2021design,tseng2021neural}, demonstrating an enhanced capability for physical data acquisition and manipulation. So far, these early endeavors have been limited to two-dimensional (2D) RGB imaging or classification problems. In this paper, we present end-to-end metaoptics inverse design for single-shot \emph{multi-channel} imaging beyond 2D RGB information: reconstruction of several depth, spectral and polarization channels \emph{simultaneously} from a single monochrome image (Section~\ref{sec:theory}). As a key result, we show that, even though demultiplexing is not a designated/prescribed objective, inverse design automatically leads to \emph{spatial demultiplexing} of the multiple channels into \emph{spontaneous domains}---distinct regions in the detected image for different channels, whose locations are not pre-designated but are optimally discovered during the course of optimization (Sections~\ref{sec:results}~and~\ref{sec:discuss}). In contrast to data-driven approaches such as neural networks~\cite{baek2020end}, our framework is physically interpretable, does not overfit despite a small generic training set, and is fully validated against ground truths vastly different from those of the training set. Specifically, we present metasurface designs for 16-color imagers with $5$--$12$\% reconstruction error (under 1\% image noise), a 4-color/4-depth imager with 5\% error, and a 2-color/2-depth/4-polarization imager with 2\% error (Section~\ref{sec:results}~and~Appendix~B). All the presented designs take into account fabrication constraints and are compatible with large-scale metasurface lithography~\cite{li2021inverse}. In practice, our method only requires a single calibration step (via measurement or calculation of the point spread function) and is amenable to arbitrary material platforms and differentiable reconstruction algorithms. Our results highlight the power of full-wave optics design with subwavelength components, whereas scalar diffractive optics could struggle to distinguish different wavelengths and polarizations due to limited dispersion and polarization sensitivity~\cite{engelberg2020advantages}. 

A major aspiration of metasurface technology has been to realize aberration-free focusing via an ultra-thin interface, directly replacing traditional bulky lenses~\cite{khorasaninejad2017visible}. While there has been significant progress towards this goal~\cite{chen2018broadband}, most metalenses suffer from fundamental space-bandwidth limits on wave focusing~\cite{presutti2020focusing}. Although nanophotonic inverse design has introduced several innovations to metaoptics architectures~\cite{molesky2018inverse,sell2017large,lin2018topology,lin2019topology,pestourie2018inverse,shi2020continuous}, further disruptive improvements await the advent of mature three-dimensional (3D) nanofabrication~\cite{lin2021computational,roques2021towards,camayd2020multifunctional}. In contrast, recent studies in end-to-end inverse design~\cite{lin2021end,burgos2021design,tseng2021neural} have unveiled ``computationally aware'' nanostructures that bear little semblance to a lens and offer capabilities beyond optics-only or computation-only designs. On the other hand, several computational techniques have been developed for retrieving depth, spectral and polarization information from a scene~\cite{pentland1987new,guo2019compact,levin2007image,greengard2006depth,monakhova2020spectral,sahoo2017single,yang2019single,rubin2019matrix}. Such techniques operate by combining multiple bulky refractive, diffractive and/or absorptive elements, often involve time-domain multiplexing (for example, scanning a scene to accumulate different shots), and typically enable the reconstruction of a single additional dimension (e.g. depth, color, \emph{or} polarization). A universal framework is still lacking, by which a \emph{single-piece} nanophotonic structure can be optimally designed to extract \emph{any and all} channels \emph{simultaneously} from a \emph{single filter-free} monochrome exposure. Our proposed end-to-end framework enables inverse design of an ultra-thin single-layer metasurface in conjunction with a simple Tikhonov-regularized reconstruction algorithm. In particular, the Tikhonov regularization is agnostic to the nature of the information channels under consideration and is thus capable of extracting any and all channels (whether they be depth, spectral, polarization, or any combination thereof).

\begin{figure}
    \centering
    \includegraphics[width=1.0\textwidth]{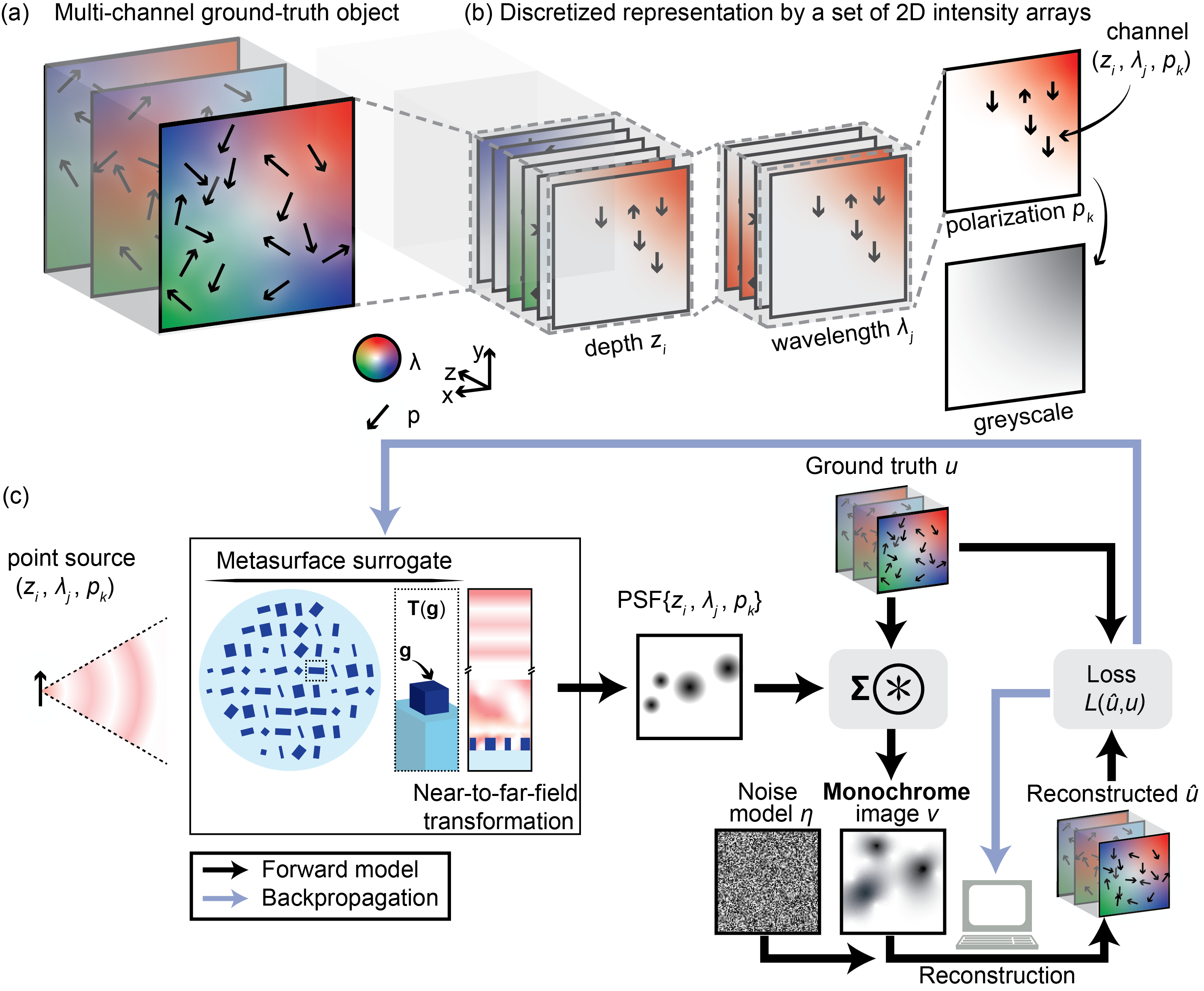}
    \caption{(a,b) A multi-channel ground truth object consists of depth, spectral and polarization channels and can be represented by a set of two-dimensional (2D) intensity arrays indexed by $(z,\lambda,p)$: a three-dimensional (3D) object can be naturally sectioned into a collection of 2D ``depth'' slices; each depth slice can be further decomposed into different ``color'' slices; each color slice is, in turn, decomposed into different ``polarization'' slices. (c) End-to-end inverse design: a metasurface frontend is optimized in conjunction with a computational reconstruction backend to minimize the reconstruction error evaluated at the end of the full pipeline.}
    \label{fig:intro}
\end{figure}

\section{Theory}
\label{sec:theory}
\subsection{Image formation model}
In conventional imaging, the optical frontend is usually modeled by an elementary phase-shift function $e^{i 2 \pi h(x,y) / \lambda}$, where $\lambda$ is the free-space wavelength and $h(x,y)$ the surface profile of the diffractive optical element~\cite{goodman2005introduction}. In nanophotonics and metaoptics, involving sub-wavelength scatterers, more detailed electromagnetic simulations are required, which must take into account richer wave effects such as multiple scattering~\cite{yu2011light,khorasaninejad2016metalenses,lin2018topology,lin2019topology,lin2019overlapping,pestourie2018inverse}. In this work, we use a Chebyshev-interpolated surrogate model ($\mathbf{T}$) under a locally periodic approximation (LPA) to efficiently simulate the transmitted electric field through a large-area metasurface~\cite{pestourie2018inverse}. Specifically, a metasurface is defined by a vector $\mathbf{g}$ characterizing the geometry of meta-atoms (such as width, height and orientation of nanopillars) while the surrogate model maps each parameter $g$ in a periodic unit cell to complex transmission coefficients. The transmitted electric field is then given by $\mathbf{E}_\text{transmitted} = \mathbf{T} (\mathbf{g}) \cdot \mathbf{E}_\text{incident}$.

In general, any ground-truth object $\mathbf{u}$ can be numerically discretized into a tensor of five dimensions including three-dimensional real space as well as color and polarization dimensions. For convenience, we denote $\mathbf{u}$ as a set of 2D $(x,y)$ intensity arrays: $\mathbf{u} \equiv \{ u_{z,\lambda,p} \}$, where each 2D array $u$ is indexed by depth ($z$), wavelength ($\lambda$) and polarization ($p$) channels (see Fig.~\ref{fig:intro}a,b). Such a ``multi-channel'' representation is naturally made for a multi-channel image-formation model, in which a single 2D monochrome image $v$ is formed by the sum of convolutions of the object channels with the corresponding point spread functions (PSFs) also indexed by $(z,\lambda,p)$:
\begin{align}
    v &= \sum_{z,\lambda,p} \mathrm{PSF}_{z,\lambda,p} \circledast u_{z,\lambda,p} + \eta, \\
    z &\in \{ z_1, z_2, ..., z_{n_z} \}, \notag \\
    \lambda &\in \{ \lambda_1, \lambda_2, ..., \lambda_{n_\lambda} \}, \notag \\
    p &\in \{ p_x, p_y, p^R_{xy}, p^I_{xy} \} \notag,
\end{align}
where $\eta$ is a generic noise term (typically modeled by zero-mean Gaussian white noise with standard deviation $\sigma$: $\eta \sim \mathcal{N}(0,\sigma^2)$~\cite{sitzmann2018end}). Note that in our model, we assume shift-invariant PSFs (valid in the paraxial regime) and only consider object intensities under incoherent illumination~\cite{goodman2005introduction}. The PSFs are computed from the surrogate model followed by near-to-far-field propagation, given a specific metasurface geometry $\mathbf{g}$. While there is no limit to the number of depths ($n_z$) or wavelength channels ($n_\lambda$), four polarization channels are sufficient to reconstruct the full Stokes vector~\cite{damask2004polarization} ($n_p \leq 4$). Those components can be understood as follows: $x$-polarized intensity channel ($p_x$), $y$-polarized intensity channel ($p_y$), the real part of the correlation between $x$ and $y$ polarizations ($ p^R_{xy}$) and the imaginary part ($ p^I_{xy} $). 

\subsection{Inverse scattering and end-to-end optimization}
Given the multi-channel image formation model, the corresponding reconstruction problem (also called inverse scattering problem) is posed as:
\begin{align}
  \min_{\{\mu_{z,\lambda,p}\}} ~ \Big\lVert v - \sum_{z,\lambda,p} \mathrm{PSF}_{z,\lambda,p} \circledast \mu_{z,\lambda,p} \Big\rVert^2 + R(\{\mu_{z,\lambda,p}\}). \label{eq:recon}
\end{align}
The reconstructed object, denoted by $\mathbf{\hat{u}} = \{ \hat{u}_{z,\lambda,p} \}$, is the solution that minimizes the problem ($\ref{eq:recon}$). Here, a regularization term $R(\cdot)$ is usually needed to make the inverse problem well-posed and well-conditioned as well as to impose any prior information such as sparsity or smoothness. A simplest choice (with minimal prior information) is the so-called Tikhonov regularization or $L_2$ norm~\cite{tarantola2005inverse} where $R(\cdot) = \alpha \lVert \cdot \rVert^2$, leading to:
\begin{align}
\mathbf{\hat{u}} = \left( \mathbf{G^TG} + \alpha \mathbf{I} \right)^{-1} \mathbf{G^Tv}, \label{eq:reconsol}
\end{align}
where, for convenience, the convolutions have been recast into a matrix notation,
\begin{align}
\mathbf{G} &= 
    \begin{bmatrix}
    \mathrm{PSF}_{z_1,\lambda_1,p_x} \circledast & \ldots & \mathrm{PSF}_{z,\lambda,p} \circledast & \ldots 
    \end{bmatrix}
\end{align}
Typically, the matrix $\mathbf{G}$ is large and dense, easily reaching over $10^5 \times 10^5$ in dimension. We use matrix-free FFT-based convolutions~\cite{goodman2005introduction,frigo2005design} in both forward and inverse scattering models to efficiently compute the action of $\mathbf{G}$ or $\mathbf{G^T}$ on arbitrary vectors without storing $\mathbf{G}$ explicitly. In particular, in Eq.~\ref{eq:reconsol}, $\mathbf{\hat{u}}$ can be obtained by the iterative conjugate-gradient method~\cite{strang2007computational}, within $\sim 100$ iterations, instead of directly computing a matrix inverse. Our end-to-end inverse design considers the entire pipeline (see Fig.~\ref{fig:intro}c) and can be formulated as minimizing the average reconstruction error:
\begin{align}
    &\min_{\mathbf{g},\alpha} \quad L(\mathbf{\hat{u}},\mathbf{u}) \stackrel{\triangle}= \langle \lVert \mathbf{u} - \mathbf{\hat{u}} \rVert^2 \rangle_{\mathbf{u},\eta} \\
    &\mathbf{\hat{u}} = \left( \mathbf{G^TG} + \alpha \mathbf{I} \right)^{-1} \mathbf{G^Tv} \notag \\
    &v = \sum_{z,\lambda,p} \mathrm{PSF}_{z,\lambda,p} \circledast u_{z,\lambda,p} + \eta \notag \\
    &\mathrm{PSF} = |\mathrm{FF}\left( \mathbf{T} (\mathbf{g}) \cdot \mathbf{E}_\text{incident} \right)|^2. \notag
\end{align}
Here, $\langle \cdot \rangle_\mathbf{u,\eta}$ denotes averaging over training data as well as image noise; the training dataset consists of a few randomly-generated ground truths (e.g., random patterns drawn from a uniform distribution). $\mathrm{FF}$ denotes the near-to-far-field propagation to the detector plane---a convolution of the transmitted electric fields with the free-space Green's function~\cite{goodman2005introduction}. In our end-to-end framework, the gradients are back-propagated through the entire pipeline all the way to the metasurface parameters, and are efficiently handled by an in-house implementation of the adjoint method~\cite{strang2007computational} (see Appendix A) instead of solely relying on popular automatic differentiation libraries~\cite{maclaurin2015autograd} which perform poorly for differentiating through iterative algorithms such as the conjugate-gradient method. The reconstruction accuracy of an optimized design is validated over vastly different ground truths (distinct from training objects).

\section{Results}
\label{sec:results}
\begin{figure}
    \centering
    \includegraphics[width=0.98\textwidth]{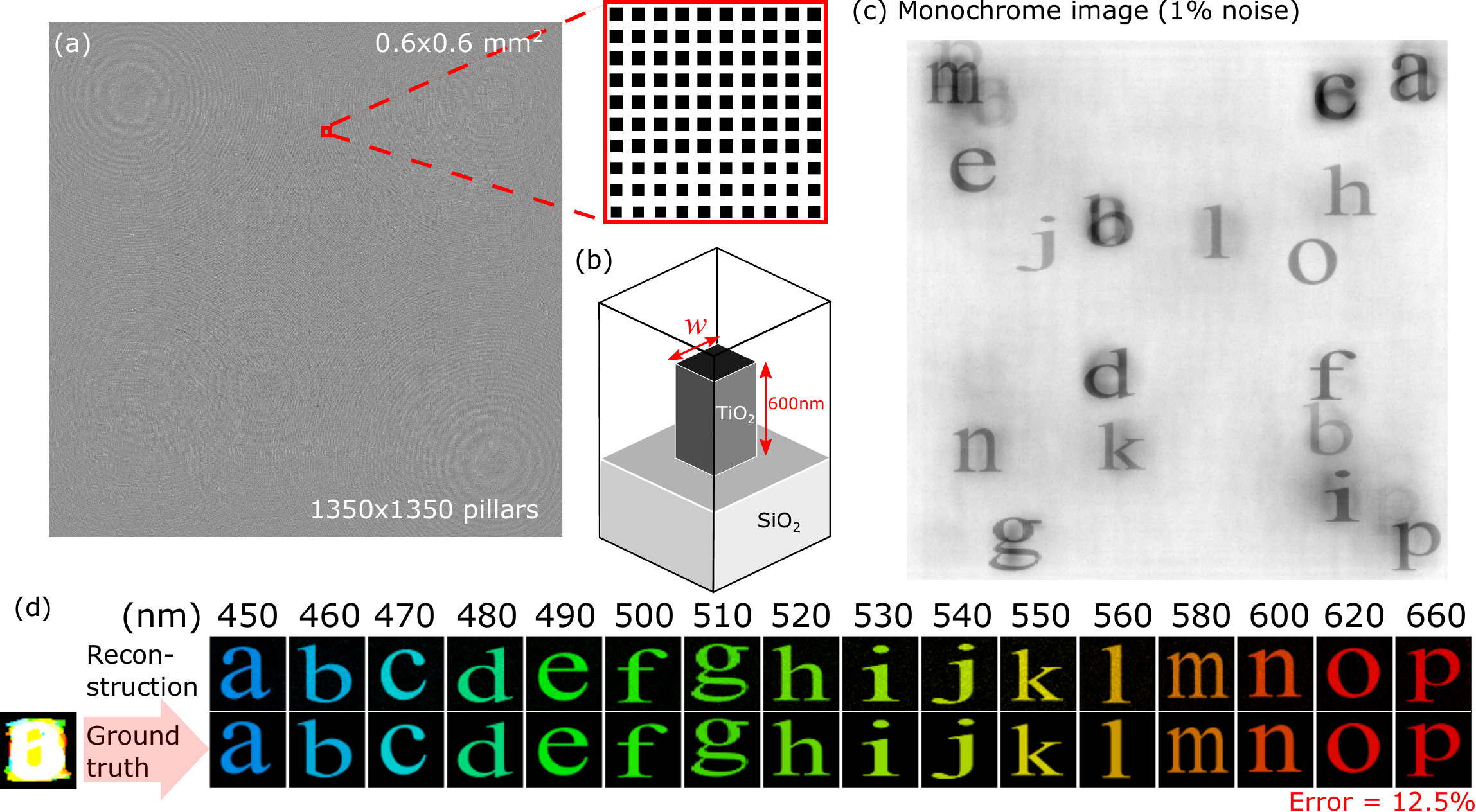}
    \caption{(a) Design of a metasurface multi-spectral imager that can reconstruct 16 color channels from 450 nm (blue) to 660 nm (red). Inset: zoom-in of the design. (b) Each unit cell has a period of $465~\mathrm{nm}$, consisting of a square nanopillar. The pillar has a height of $600~\mathrm{nm}$ and a width of $60~\mathrm{nm} \leq w \leq 405~\mathrm{nm}$. (c) Monochrome image of the synthetic object shown in (d). The wavelengths are spatially demultiplexed onto distinct domains on the single-shot monochrome image, captured 1 mm away from the metasurface by a CCD array of $400\times 400$ pixels (each pixel has an area of $1.4 \times 1.4~\mathrm{\mu m}^2$). (d) Reconstruction of a synthetic ground truth---a multi-spectral picture of letters `a' to `p', situated 2~cm away from the metasurface and each letter emitting a different wavelength. Note that the letters in the ground truth cannot be distinguished by the naked eye (inset on the left of the ground truth row). Computationally, the ground truth is represented by a set of 16 intensity arrays, each of which is a $50\times 50$-pixel image of a letter (with $25~\mathrm{\mu m}$ resolution).  The ground truth and the reconstruction are color-coded for a visual interpretation of the wavelengths. The reconstruction error is $12.5\%$ under $1\%$ image noise.}
    \label{fig:16color}
\end{figure}

We now show how our framework can be utilized to inverse-design metaoptics with multi-channel reconstruction capability (depth, spectral, and polarization). We denote the dimensions of a ground truth object as $n_\text{ch} \times m\times m$---a set of $n_\text{ch}$ arrays each with $m \times m$ pixels (note that $n_\text{ch}=n_z n_\lambda n_p$), while the monochrome image is a \emph{single} 2D array of $n \times n$ pixels. In this work, we choose $n^2 \geq n_\text{ch} m^2$, that is, there are at least as many image pixels as the \emph{total} size of the object---an over-determined inverse problem, suitable for Tikhonov regularization which harbors minimal assumptions about the nature of the object. 

First, we design a 16-color metasurface imager made up of $600~\mathrm{nm}$-tall TiO$_2$ pillars on silica (Fig.~\ref{fig:16color}a,b)---a design platform compatible with large-area lithographic fabrication as recently demonstrated in millimeter-scale achromatic metasurfaces~\cite{li2021inverse}. A Chebyshev-interpolated surrogate model maps the width of each pillar inside a unit cell (465 nm period) to transmission coefficients at 16 different wavelengths across the visible spectrum ($450 - 660~\mathrm{nm}$). The \emph{single-shot monochrome} image shows \emph{spatial demultiplexing} of the wavelength channels (Fig.~\ref{fig:16color}c). Interestingly, the imager does not solely rely on the demultiplexing effect; for example, there is channel replication, e.g. $\lambda_2$ (460 nm) channel, and a small degree of hybridization, e.g. between $\lambda_1$ (450 nm) and $\lambda_2$ (460 nm) channels. While the human eye is not equipped to recover all the information encoded in hybridization and redundancy, these apparent ``imperfections'' do not necessarily represent information loss. In particular, the apparent mixing between different channels does not preclude information recovery since the channels can be readily reconstructed by computation (as long as the corresponding PSFs are distinctly non-degenerate), leading to a reconstruction error of $12.5\%$ under $1\%$ Gaussian image noise (Fig.~\ref{fig:16color}e). We note that a signal-to-noise ratio of $\sim 100$ (1\% noise) can be readily achieved by modern electronic sensors~\cite{sitzmann2018end}. Furthermore, the apparent residual fine-grain noise in the reconstruction of non-random objects can be easily removed by simple de-noising routines. 

In this example, we considered a simple geometry (a square pillar) suited for photo-lithographic mass production; utilizing a more complex geometry, such as a holey pillar, allowing for more degrees of freedom to manipulate incident wavefronts, leads to even better performance (5\% error with 1\% noise, see Appendix B). Our methods are also amenable to inverse-design techniques allowing for freeform geometries, involving domain decomposition methods with larger unit cells and full topology optimization~\cite{molesky2018inverse,lin2019overlapping}. We emphasize that our framework does not seek a ``heavily-processed imitation'' of the ground truth; it looks for a faithful reconstruction which is \emph{stable} under moderate noise, and should be applicable for imaging \emph{any} object, including random ones (see Appendix B). If desired, additional processing may be used, such as convolutional neural networks, which can be trained to ``interpolate'' a particular distribution of objects, enhance the reconstruction of \emph{that} class of objects, perform image segmentation, or classification on the reconstructed objects.

% \begin{figure}
%     \centering
%     \includegraphics[width=0.98\textwidth]{specimgs.pdf}
%     \caption{Additional spectral imager designs: 3 colors (upper), 9 colors (middle), 16 colors (lower). A color channel has $m \times m$ pixels while the monochrome image has $n \times n$ pixels. The image is corrupted by 1\% noise.}
%     \label{fig:specimgs}
% \end{figure}

\begin{figure}
    \centering
    \includegraphics[width=0.98\textwidth]{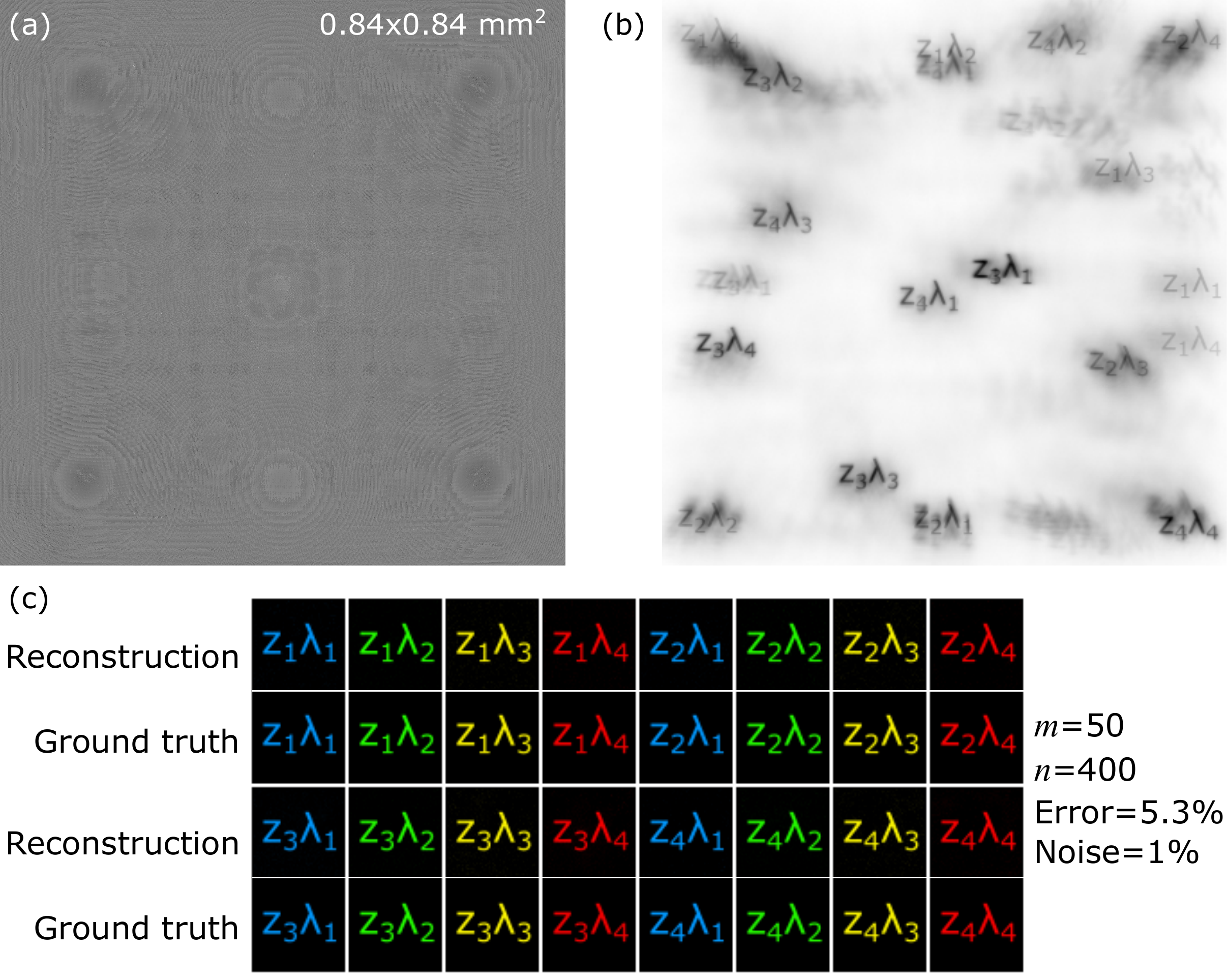}
    \caption{Depth-spectral imager. (a) Metasurface depth-spectral imager design. (b) Monochrome image of a synthetic multi-dimensional object consisting of 4 depths $\times$ 4 color channels. The channels in the test object are artificially synthesized as $m \times m$-pixel images of the channel indices ($m=50$). The monochrome image has $n \times n$ pixels ($n=400$) and is corrupted by 1\% noise, leading to (c) a reconstruction error of 5.3\%. Note that $z_i \in \{2,4,6,8\}~\mathrm{cm},~\lambda_j \in \{ 470,520,582,660\}~\mathrm{nm}$.}
    \label{fig:depspec}
\end{figure}

\begin{figure}
    \centering
    \includegraphics[width=0.98\textwidth]{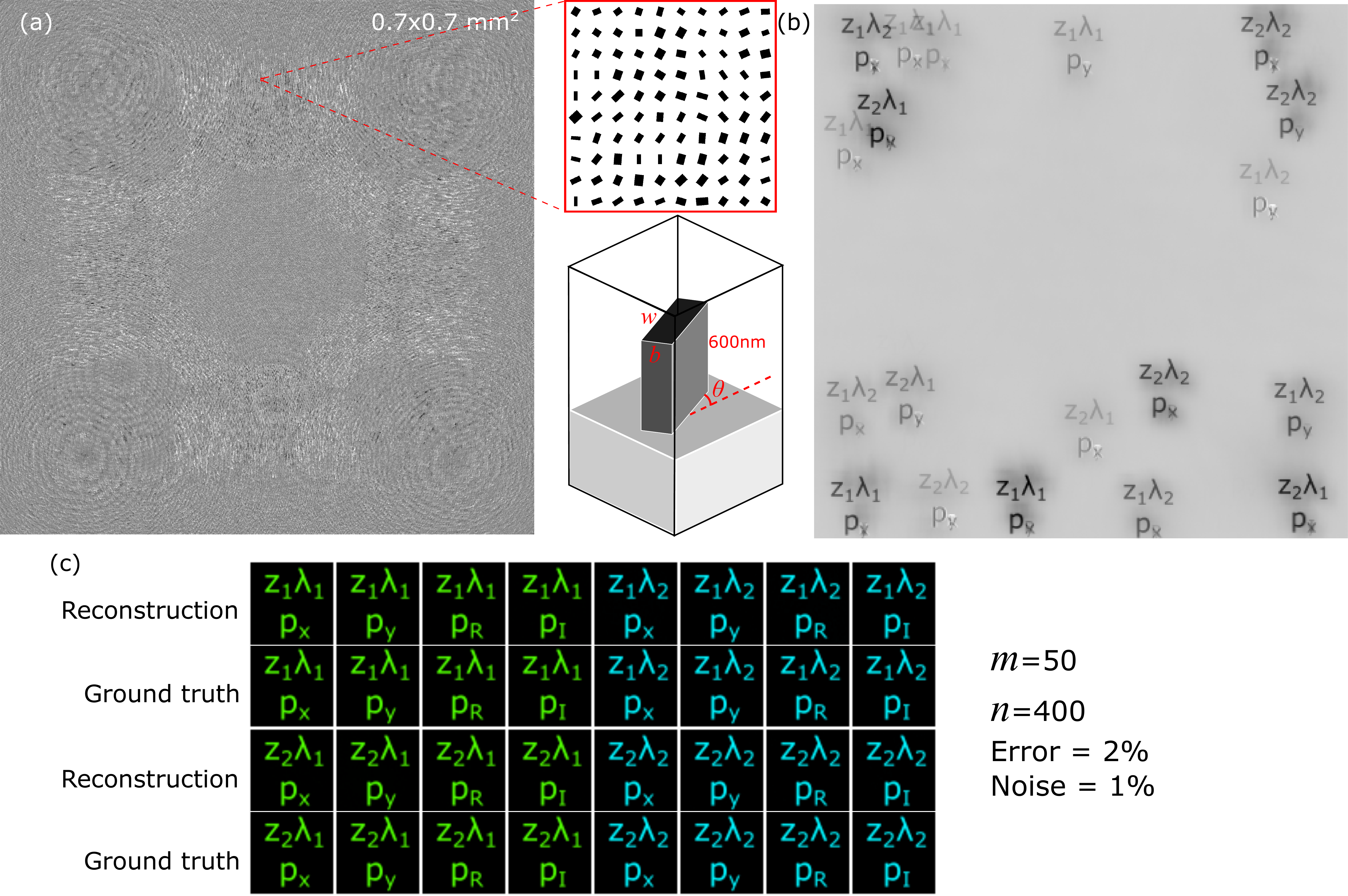}
    \caption{Spectro-polarimetric-depth imager. (a) The metasurface consists of TiO$_2$ nanopillars, each characterized by width ($w$), breadth ($b$) and orientation angle ($\theta$), where $60~\mathrm{nm}\leq w,b \leq 299~\mathrm{nm}$. (b) Monochrome image of a synthetic multi-dimensional object consisting of 2 depths $\times$ 2 colors $\times$ 4 polarization channels. The channels in the test object are artificially synthesized as pictures of the channel indices with $m \times m$ pixels ($m=50$). The monochrome image has $n \times n$ pixels ($n=400$) and is corrupted by 1\% noise, leading to (c) a reconstruction error of 2\%. Note that $z_i \in \{1.7,3.4\}~\mathrm{cm},~\lambda_j \in \{ 532,488\}~\mathrm{nm}$.}
    \label{fig:spd}
\end{figure}

Apart from spectral imagers, our framework is powerful in that it is straightforward to extract \emph{any and all} kinds of channels. For example, we design a depth-spectral imager (Fig.~\ref{fig:depspec}) that can reconstruct 4 depth channels $\times$ 4 wavelength channels. Additionally, as a proof of concept, we also design an ``all-in-one'' imager (Fig.~\ref{fig:spd}) that can reconstruct 2 depth channels $\times$ 2 wavelength channels $\times$ 4 polarization channels. In that case, spontaneous spatial demultiplexing discovered via inverse design is observed for channels that are a combination of a given depth and polarization (i.e. channels sharing the same depth but having different polarizations are also demultiplexed, and vice-versa). On the other hand, a greater degree of hybridization is seen to arise in between the depth channels. This originates from the limited geometric control of the \emph{local} metasurface design we have chosen, which cannot provide sufficiently strong \emph{spatial} dispersion to fully separate the depth channels. In future works, we will engineer larger unit cells, higher diffraction orders, and cascaded metamaterials to induce strongly \emph{non-local}, spatially-dispersive effects~\cite{lin2018topology,lin2021computational}.

\section{Discussion and Outlook}
\label{sec:discuss}
The central result of this work is the realization of metaimaging based on \emph{spontaneous} demultiplexing of multi-channel information into distinct spatial domains, whose locations appear irregular but are optimally determined by end-to-end inverse design. This is in contrast to the situation where such domains would be dictated by a user, as would be the case for conventional optics-only designs such as color splitters~\cite{presutti2020focusing}, or even hybrid systems~\cite{rubin2019matrix}. The end-to-end automated discovery naturally leads to an optimal demultiplexing scheme with minimal hybridization between channels; in contrast, a human-designated scheme, such as a regular lattice of focal spots, may not be permitted by the available degrees of freedom in the metasurface, resulting in noisy crosstalk,  sub-optimal PSFs, and poor resolution. Intuitively, the demultiplexing effect is enabled by the Tikhonov reconstruction backend, which does not attempt to learn from the specific training data, but ``judiciously'' nudges the optical frontend to separate the incoming channels in order to reduce the reconstruction error. Therefore, our method is physically interpretable and data-efficient. It only requires a small training set, for example, as few as 30 training data drawn from a uniform distribution in the case of the 16-color imager (Fig.~\ref{fig:16color}). At the same time, the final optimized designs achieve robust reconstruction performance with consistent accuracy over vastly different sets of ground truths (whether they are pictures like letters or patterns like random dots, see Appendix B). This is in contrast to recent works~\cite{baek2020end} using phase masks and neural networks, which require large, diverse and carefully curated training sets and do not lead to spatial demultiplexing. 

One conceivable limitation of our multi-channel imagers is that the transverse dimensions of the object must be significantly smaller than those of the detector; therefore, the device is not suitable for reconstructing the entire natural field of view corresponding to the size of the detector. In practice, a narrower operational field of view may be realized by an appropriate aperture, a directed flash, or by selective illumination (a common technique in microscopy)~\cite{saleh2019fundamentals,mertz2019introduction}. The field of view can be enlarged by designing larger-area metasurfaces or by taking into account out-of-field-of-view light in the image-formation model. On the other hand, the inverse problem is under-determined if we choose the same transverse dimensions for the object and the detector. Such a problem requires additional priors on the object, and a regularization scheme like Tikhonov may not be sufficient. One powerful prior in image processing is sparsity, and a theoretically rigorous technique for reconstructing sparse objects is called compressed sensing~\cite{donoho2006compressed}. In another manuscript under preparation, we will present a fully end-to-end inverse design framework with a compressed-sensing backend.  Ultimately, future backends may be realized by new architectures that combine classical algorithms (such as Tikhonov and CS, which are theoretically rigorous and physically interpretable) and deep neural networks (which are best suited for learning deep data priors). Moreover, the performance of ultra-compact nanophotonic devices (such as depth and spectral sensitivities) can be further enhanced if we transcend the limitations imposed by LPA and expand the available degrees of freedom to encompass the full Maxwell physics. In future work, we hope to explore end-to-end inverse design using more sophisticated domain decomposition methods~\cite{lin2019overlapping} and scattering-matrix formulations~\cite{benzaouia2021quasi}, or by cascading non-local metamaterials and 3D photonic crystals with local metasurfaces.

\section*{Funding}
Z.~Lin, C.R.C., R.P., M.S. and S.G.J. were supported in part by the U.~S.~Army Research Office through the Institute for Soldier Nanotechnologies under award number W911NF-18-2-0048. Z.~Lin and R.P. were partially supported by the MIT-IBM Watson AI Laboratory under Challenge 2415. Z.~Li and F.C. were supported by MURI AFOSR grant  FA9550-21-1-0312.

%\bibliographystyle{unsrt}
%\bibliography{ref}

%%%%%%%%%%%%%%

%%%%%%%%%%%%%

\appendix
\renewcommand\thefigure{A.\arabic{figure}}    
\setcounter{figure}{0}

\section{Adjoint gradient}
The reconstruction $\mathbf{\hat{u}}$ under $\eta = 0,~\alpha>0$ is obtained by iteratively solving the equation:
\begin{align}
    \left( \mathbf{G^TG} + \alpha \mathbf{I} \right) \mathbf{\hat{u}} = \mathbf{G^TGu}, 
\end{align}
using the conjugate-gradient method without forming an explicit matrix. Note that the transpose of a convolution kernel is a convolution with the mirror image of the original kernel. Therefore, $\mathbf{G^T}$ is simply convoluting with mirrored PSFs and then vertically stacking the results (in order to have output of the same size and shape as $\mathbf{u}$.)

Given a function $f(\mathbf{\hat{u}}(\mathbf{g}))$, we outline how to use the adjoint method~\cite{strang2007computational} to find $\frac{\partial f}{\partial \mathbf{g}}$. 
\begin{align}
     \frac{\partial f}{\partial \mathbf{g}} &= \frac{\partial f}{\partial \mathbf{\hat{u}}} \cdot \frac{\partial \mathbf{\hat{u}}} {\partial \mathbf{g}} \\
     &= \mathbf{\Lambda} ~\cdot \frac{\partial \mathbf{G^TG}}{\partial \mathbf{g}} \left( \mathbf{u} - \mathbf{\hat{u}} \right)
\end{align}
where the adjoint variable $\mathbf{\Lambda}$ is given by
\begin{align}
    \left( \mathbf{G^TG} + \alpha \mathbf{I} \right) \mathbf{\Lambda} &= \frac{\partial f}{\partial \mathbf{\hat{u}}}
\end{align}
We may find $\mathbf{\Lambda}$ using the same iterative solver that we used to find $\mathbf{\hat{u}}$.

The trickier issue is to find $ \frac{\partial \mathbf{G^TG}}{\partial \mathbf{g}}$. If one carries out the algebra faithfully, one may find that the inner product sandwiching the tricky derivative boils down to a cross-correlation between $\mathbf{\Lambda}$ and $\mathbf{G}\left( \mathbf{u} - \mathbf{\hat{u}} \right)$. However, we can exploit the \texttt{autograd} automatic-differentiation (AD) package~\cite{maclaurin2015autograd} in Python to compute this derivative effortlessly as follows (pseudo-code):
\begin{verbatim}
    def innerdv(x,a,b):
        
        def aGTGb(x):
            ... # compute and return aGTGb given design parameters x 
            ... # by ``autograd-able'' convolutions
            
        g = autograd.grad(aGTGb)
        
        return g(x)
\end{verbatim}
The desired product $\mathbf{\Lambda} ~\cdot \frac{\partial \mathbf{G^TG}}{\partial \mathbf{g}} \left( \mathbf{u} - \mathbf{\hat{u}} \right)$ is then simply given by \texttt{innerdv(g,Lambda,u-uhat)}.

\newpage

\section{Metasurface with holey pillars}
\begin{figure}[h!]
    \centering
    \includegraphics[scale=0.40]{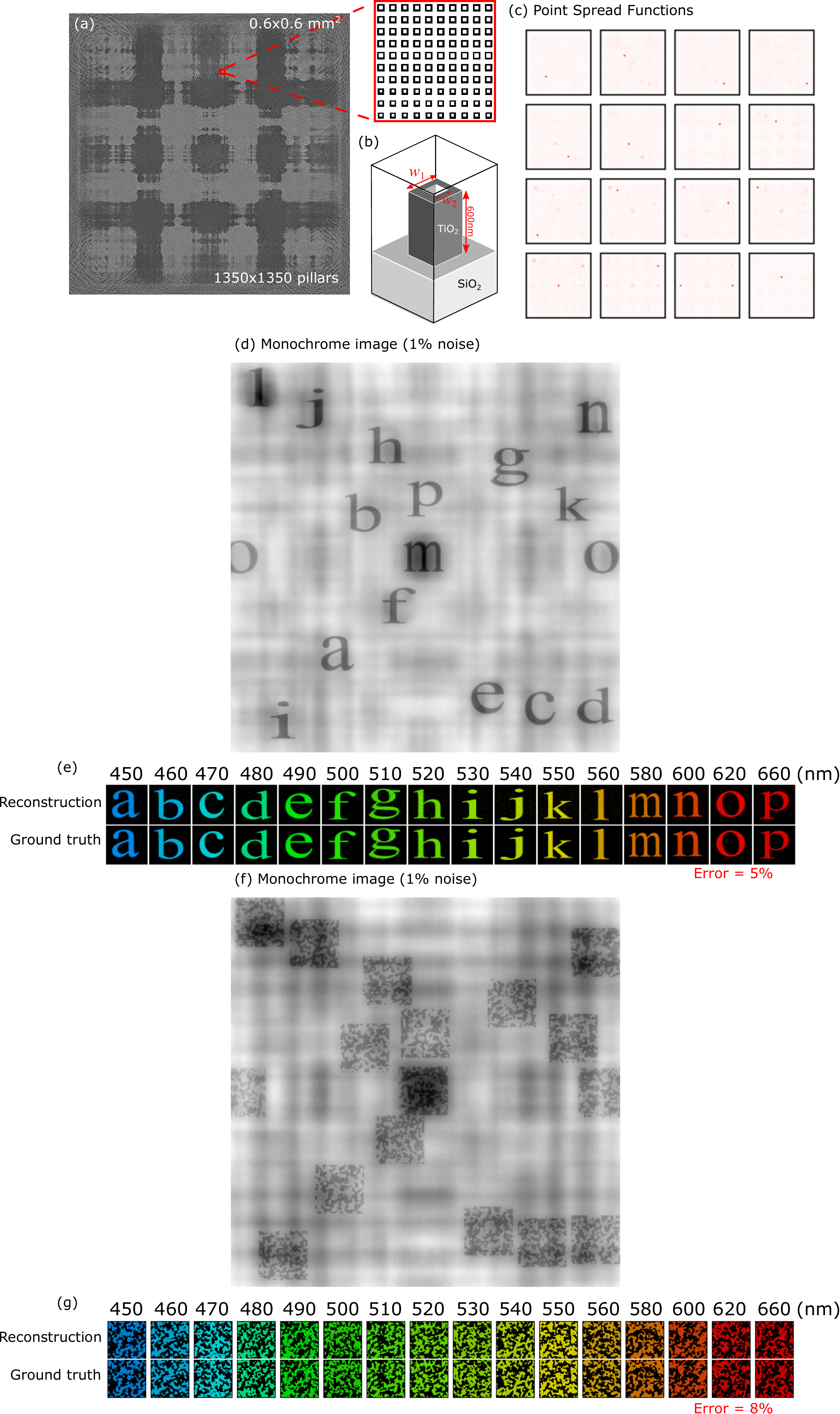}
    \caption{(a-c) A 16-color imager with ``holey pillars'' and the PSFs. The metasurface has an average transmission efficiency of > 55\%. It can accurately reconstruct colored letters (d,e) as well as a random ground truth (f,g).}
    \label{fig:hoop}
\end{figure}
\end{document}